\documentclass[aps,prb,groupedaddress,twocolumn]{revtex4-2}
\usepackage{amsmath}
\usepackage{graphicx}
\usepackage{siunitx}

\begin{document}

\title{Spin-dependent coupling of supercurrent and nonequilibrium quasiparticles in high-field superconductors}


\author{P. Maier}
\author{D. Beckmann}
\email{detlef.beckmann@kit.edu}
\affiliation{Institute for Quantum Materials and Technologies, Karlsruhe Institute of Technology (KIT), D-76021 Karlsruhe, Germany}


\date{\today}

\begin{abstract}
We report on an experimental investigation of the combined effect of nonequilibrium quasiparticle injection and supercurrent in superconducting aluminum wires. At low temperature, we observe the supercurrent-induced coupling of energy and charge imbalance with spectral resolution. At high magnetic fields, in the presence of a Zeeman splitting of the density of states, we find evidence for an additional spin-dependent coupling which has been recently predicted theoretically.
\end{abstract}


\maketitle

\section{Introduction}

Nonequilibrium transport in spin-degenerate superconductors has been investigated intensely in the 1970s and 80s \cite{langenberg1986}. In the spin-degenerate case, the nonequilibrium distribution function is characterized by the two-fold particle-hole degree of freedom, described by a ``longitudinal'' energy and a ``transverse'' charge mode. Quasiparticles are coupled to the superconducting condensate, and one of the most striking implications is a conversion between energy and charge modes induced by a supercurrent  \cite{schmid1975,schmid1979,pethick1979b,clarke1980}. The energy-charge conversion can be understood in terms of the Doppler shift of the quasiparticle spectrum due to the superfluid velocity. Experimentally, the conversion was observed by applying a temperature gradient and a supercurrent simultaneously to a superconducting wire \cite{clarke1979,fjordboge1981,heidel1981}.

Recently, the field of nonequilibrium superconductivity has been reinvogorated by the investigation of spin-polarized quasiparticle transport \cite{johnson1994,poli2008,yang2010,wakamura2015,beckmann2016}. The two-fold spin degree of freedom leads to additional spin and spin-energy nonequilibrium modes \cite{morten2004}.
Part of the motivation for these investigations comes from the idea of using spin to implement electronic functionality in the context of superconducting spintronics \cite{linder2015,eschrig2015}, either via spin-polarized supercurrents or nonequilibrium quasiparticles. For example, spin-polarized quasiparticles can control spin-polarized supercurrents \cite{bobkova2010,bobkov2011}, and supercurrents can control spin-polarized distributions \cite{amundsen2020}.

In addition to spin-dependent distribution functions, thin superconducting films in high magnetic fields have spin-dependent spectral properties \cite{meservey1970}. The spin splitting of the density of states leads to long-range spin transport \cite{huebler2012b,quay2013,silaev2015,krishtop2015,bobkova2015a,beckmann2016,bergeret2018,heikkila2019,kuzmanovic2020} and large spin-dependent thermoelectric effects \cite{machon2013,ozaeta2014,kolenda2016,heidrich2019}. Recently, it has been predicted that the spin-dependence of the spectral supercurrent creates an additional coupling term between supercurrent and quasiparticles in high-field superconductors, leading to conversion between spin-degenerate and spin-polarized nonequilibrium modes \cite{aikebaier2018}. 
Here, we report the experimental observation of this additional coupling term via the conversion of energy nonequilibrium to charge and spin-energy modes in high-field superconducting aluminum wires. 

\section{Experiment}
\begin{figure}
    \includegraphics[width=\columnwidth]{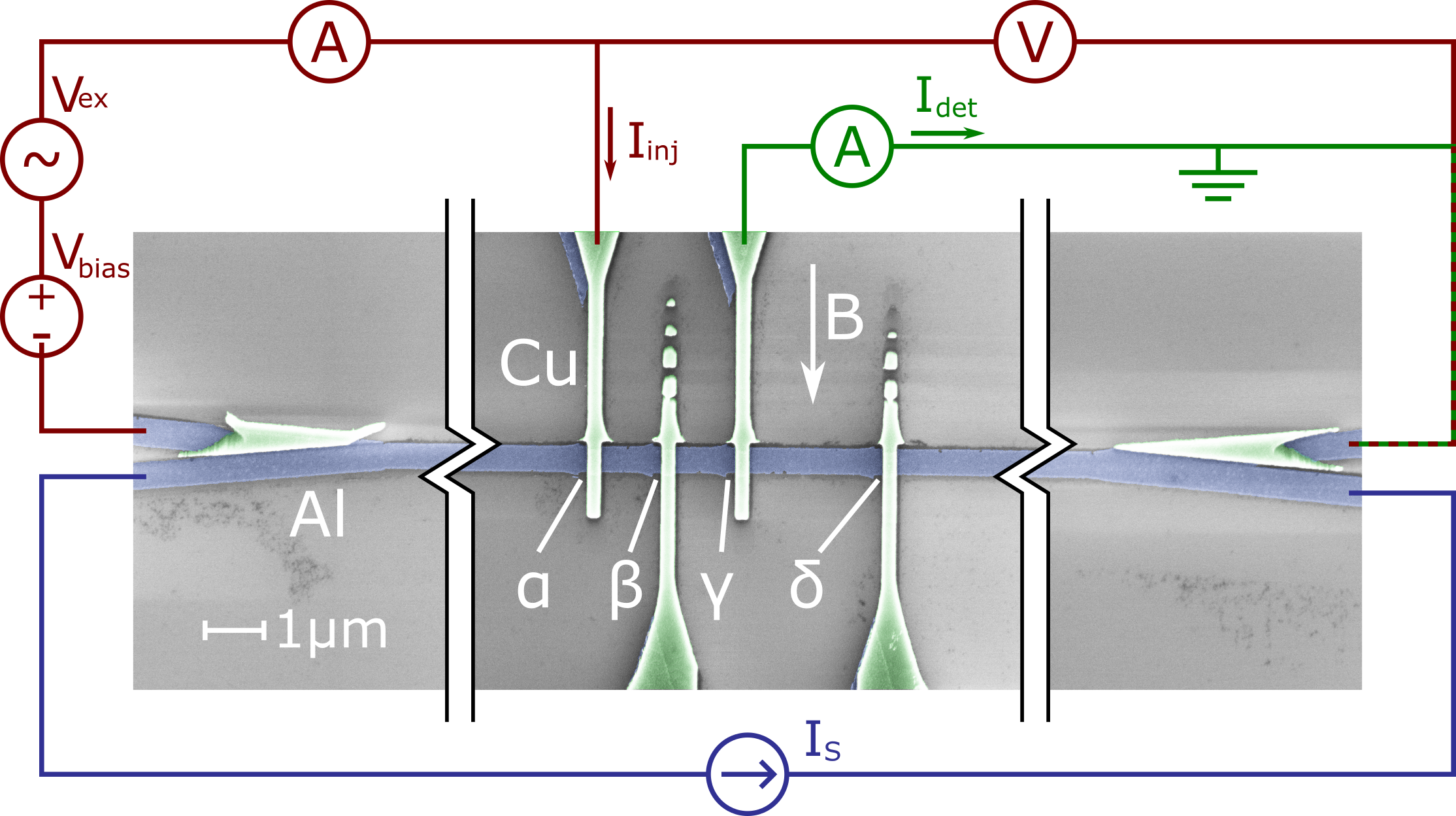}
    \caption{False-color scanning electron microscopy image of sample B. All samples consist of an aluminum strip with several copper wires attached by tunnel contacts. Contact $\alpha$ at the center of the strip is used as the injector, the remaining ones as detectors. The ends of the aluminum strip are split so that a supercurrent $I_\mathrm{S}$ can be applied without affecting the conductance measurement. The image is shortened, the total length of the aluminum strip between the splits is about \SI{24}{\micro\meter}.}
    \label{fig:sampleSEM}
\end{figure}

Nonequilibrium quasiparticle transport was investigated in two samples of similar design (labeled A and B). Figure \ref{fig:sampleSEM} shows a false-color scanning electron microscopy (SEM) image of sample B along with the measurement scheme. All samples were fabricated by electron beam lithography and shadow evaporation. The substrates are pieces of silicon wafer with \SI{1}{\micro\meter} silicon oxide. The structures consist of a long (24 - \SI{50}{\micro\meter}), thin (12 - \SI{17}{\nano\meter}) aluminum strip with split end sections and tunnel contacts with aluminum oxide barriers and copper electrodes. The normal state tunnel resistances are in the range of 1.5 - \SI{4}{\kilo\ohm}. 
The tunnel contacts are arranged with one in the center of the strip ($\alpha$) for injection and the others to one side as detectors. The small copper artifact left in the split region is separated from the superconductor by the same aluminum oxide barrier as the contacts, and therefore does not affect the measurement. 

The measurement setup consists of the local circuit ($I_\mathrm{inj}$), the nonlocal circuit ($I_\mathrm{det}$) plus the supercurrent circuit ($I_\mathrm{S}$). The local (injector) circuit was used to measure the tunnel conductance spectra via low-frequency lock-in detection with a small ac excitation $V_\mathrm{ex}$ superimposed on a dc voltage $V_\mathrm{bias}$. The local conductance was measured in a three point configuration due to limitations of the cryostat wiring, and the effect of the injector lead resistance was corrected during data analysis. The nonlocal current $I_\mathrm{det}$ due to nonequilibrium quasiparticle injection was measured simultaneously with the local conductance. In addition, a supercurrent $I_\mathrm{S}$ could be passed through the wire using the split end sections of the wire without disturbing the ac conductance measurements. 

Similar results were obtained on both samples. All data shown here were taken on sample A, except where otherwise noted.
The measurements were performed with excitation voltage rms amplitudes of about \SI{15}{\micro\volt} (Sample A) and \SI{6}{\micro\volt} (Sample B). Measurements were performed at \SI{100}{\milli\kelvin} on sample A and at \SI{20}{\milli\kelvin} on sample B if not stated otherwise. A magnetic field could be applied in plane along the direction of the copper electrodes. 

\section{Theory}
\begin{figure}
         \includegraphics[width=\columnwidth]{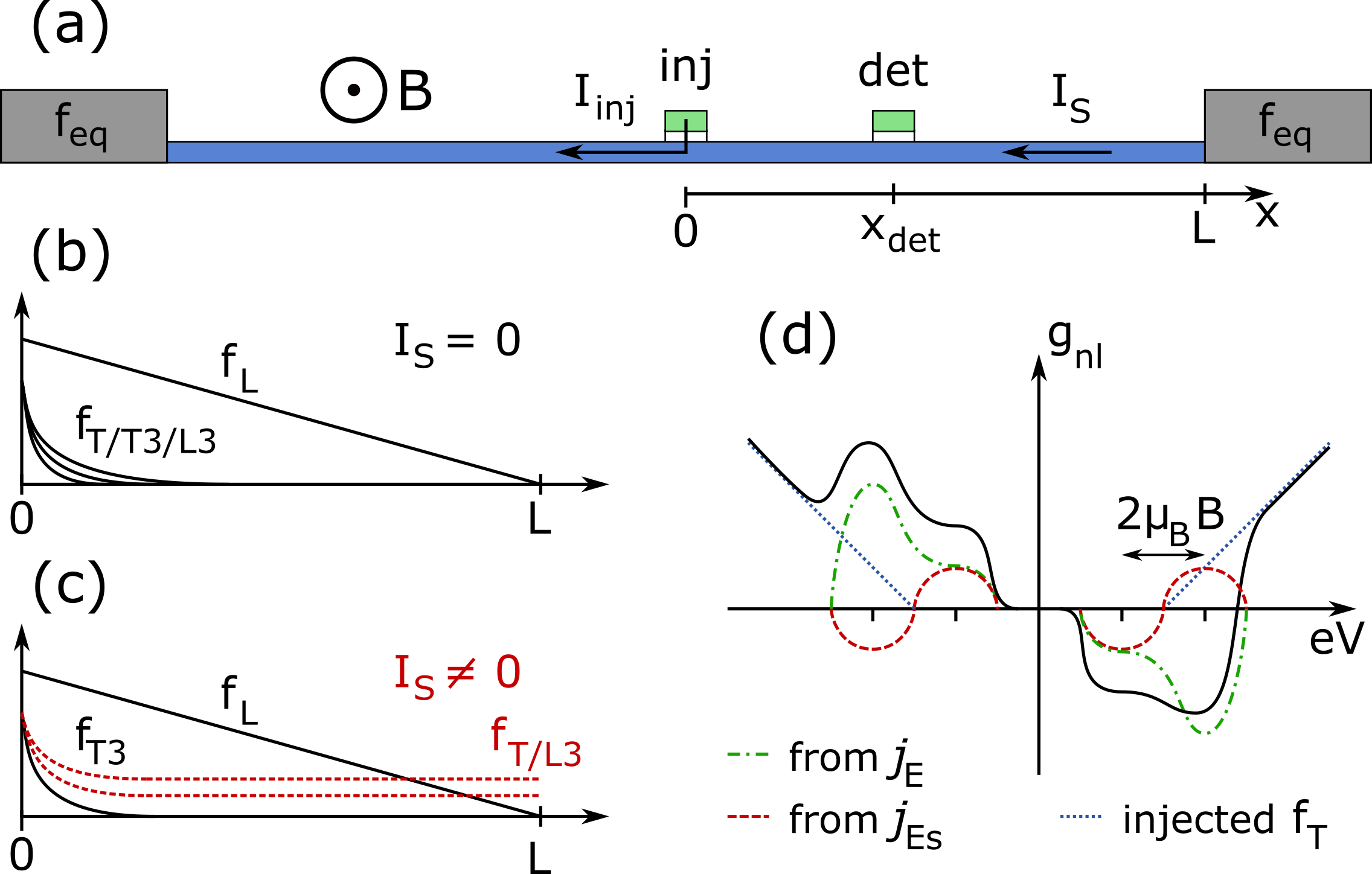}
\caption{Overview of the model. (a) Sketch of the sample geometry. A one-dimensional superconducting wire is placed between two equilibrium reservoirs. Quasiparticles are injected via a tunnel junction at $x=0$, and detected via a second junction at $x=x_\mathrm{det}$. In addition, a supercurrent $I_\mathrm{S}$ can be passed through the wire. (b) Nonequilibrium modes without supercurrent. The $f_\mathrm{L}$ mode falls linearly, all other modes decay rapidly. (c) Nonequilibrium modes with supercurrent. Coupling to the supercurrent generates $f_\mathrm{T}$ and $f_\mathrm{L3}$ proportional to the constant gradient of $f_\mathrm{L}$. (d) Nonlocal conductance contributions. Without supercurrent, the injected $f_\mathrm{T}$ mode creates a symmetric contribution (dotted line). The supercurrent coupling terms $j_\mathrm{E}$ and $j_\mathrm{Es}$ generate antisymmetric contributions. These contributions have equal/opposite sign in the lower/upper Zeeman band.}
    \label{fig:sampleModel}
\end{figure}

In this section, we give a simplified description of the theory of our experiment, focussing on the features relevant to understand the experimental results. The full model used for the numerical simulations is given in the appendix. 

We are mostly interested in the behavior at high magnetic fields, where the quasiparticle energies acquire a Zeeman splitting $2\mu_\mathrm{B}B$. As a consequence, all spectral properties are spin-dependent, and can be conveniently decomposed into a spin-symmetric and spin-antisymmetric part. For example, the spin-resolved density of states $N_\downarrow(E)$ and $N_\uparrow(E)$ can be decomposed into $N_\pm(E)=(N_\downarrow\pm N_\uparrow)/2$, where $E$ is the quasiparticle energy.

Figure \ref{fig:sampleModel}(a) shows a sketch of the model geometry. The sample is modeled as a quasi-onedimensional wire along the $x$-axis, terminated at both ends by equilibrium reservoirs. An injector junction is placed at the center ($x=0$), and a detector junction is placed at $x=x_\mathrm{det}$. The total length of the wire is $2L$. The magnetic field $B$ is applied in-plane, perpendicular to the wire.

The nonequilibrium state of a spin-split dirty-limit superconductor can be described by four distribution functions, the nonequilibrium modes $f_\mathrm{L}$, $f_\mathrm{T3}$, $f_\mathrm{T}$ and $f_\mathrm{L3}$. They describe energy, spin, charge, and spin-energy imbalance of the quasiparticle excitations, respectively  \cite{morten2004,silaev2015,bobkova2015a}. Nonequilibrium in our experiment is driven by tunnel injection. Figure \ref{fig:sampleModel}(b) shows the qualitative behavior of the four nonequilibrium modes without applied supercurrent. The charge and spin-dependent modes $f_\mathrm{T}$, $f_\mathrm{T3}$ and $f_\mathrm{L3}$ decay relatively fast due to charge relaxation or spin flips. The charge relaxation length is a few $\mathrm{\mu m}$ at zero field in our structures, but drops very quickly upon increasing the magnetic field \cite{huebler2010,huebler2012b,wolf2013} due to orbital depairing. The spin relaxation length is typically a few hundred nm in our structures \cite{huebler2012b,wolf2014c}, smaller than the contact spacing of the present experiment. The $f_\mathrm{L}$ mode relaxes only via inelastic scattering, which is weak at the low temperatures of our experiment. The electron-phonon relaxation length is typically a few 100 $\mathrm{\mu m}$ in metal wires at temperature far below $1~\mathrm{K}$ \cite{giazotto2006}, much larger than the length of our wires. Electron-electron scattering does not relax energy, but leads to a thermalization of the nonequilibrium distribution. Previous comparison of theory to similar experiments on high-field superconductors have shown that neglecting inelastic scattering is a reasonable assumption \cite{silaev2015,heidrich2019}, with at most small deviations due to thermalization by electron-electron scattering \cite{heidrich2019}. We therefore neglect inelastic scattering in the model.

Due to the weak relaxation, $f_\mathrm{L}$ is the dominant mode created by tunnel injection. Neglecting all other modes, $f_\mathrm{L}$ is given by
\begin{equation}
f_\mathrm{L}(x)=G_\mathrm{inj}R\frac{N_+f_\mathrm{L}^\mathrm{inj}(V_\mathrm{inj})}{D_\mathrm{L}+G_\mathrm{inj}RN_+}\left(1-\frac{x}{L}\right),
\label{eqn:fL}
\end{equation}
where $G_\mathrm{inj}$ is the normal-state injector conductance, $R$ is the normal-state resistance of the left and right branches of the superconducting wire in parallel, $D_\mathrm{L}$ is the spectral diffusion constant of the $f_\mathrm{L}$ mode, and $f_\mathrm{L}^\mathrm{inj}(V_\mathrm{inj})$ is the injector distribution function. $f_\mathrm{L}$ falls linearly towards the ends of the aluminum strip. Note that Fig.~\ref{fig:sampleModel}(b) is not to scale, but only illustrates the spatial dependence qualitatively. Actually, $f_\mathrm{L}$ is orders of magnitude larger than the other modes.

When a supercurrent is applied to the wire, transport of all nonequilibrium modes is coupled. Far from the injector, the part of the kinetic equation relevant for our experiment is
\begin{equation}
    \begin{pmatrix}
    R_\mathrm{T} & R_\mathrm{L3}\\
    R_\mathrm{L3} & R_\mathrm{T}+S_\mathrm{L3}
    \end{pmatrix}
    \begin{pmatrix}f_\mathrm{T}\\ f_\mathrm{L3}\end{pmatrix}
    =
\begin{pmatrix} j_\mathrm{E}\nabla\phi\\ j_\mathrm{Es}\nabla\phi\end{pmatrix} \nabla f_\mathrm{L}.
\end{equation}
$j_\mathrm{E}$ and $j_\mathrm{Es}$ are the spin-symmetric and antisymmetric parts of the spectral supercurrent and $\nabla\phi$ is the superconducting phase gradient. $R_\mathrm{T}$ and $R_\mathrm{L3}$ describe charge relaxation, and $S_\mathrm{L3}$ is the spin relaxation rate. The historic experiments correspond to $B=0$, where $j_\mathrm{Es}$ and $R_\mathrm{L3}$ are zero and only $f_\mathrm{T}$ is generated. Thus, we arrive at the qualitative picture that tunnel injection drives $f_\mathrm{L}$, and then the gradient of $f_\mathrm{L}$ in combination with the supercurrent generates $f_\mathrm{T}$ and $f_\mathrm{L3}$ along the wire. The generation is balanced by charge and spin relaxation. The qualitative behavior is shown in Fig.~\ref{fig:sampleModel}(c). For the constant $\nabla f_\mathrm{L}$ following from the linear decay in Eq.~(\ref{eqn:fL}), the generated modes are independent of position. We would also like to note that the generated modes are proportional to $\nabla \phi$, and therefore odd functions of supercurrent. This property will be used later in the data analysis.

The nonlocal differential conductance $g_\mathrm{nl}=dI_\mathrm{det}/dV_\mathrm{inj}$ depends on the $f_\mathrm{T}$ and $f_\mathrm{L3}$ modes at the detector contact via
\begin{equation}
I_{\mathrm{det}}=-\frac{G_\mathrm{det}}{2e}\int^{\infty}_{-\infty}dE(N_{+}f_\mathrm{T}+N_{-}f_\mathrm{L3}).
\end{equation}
Here, $G_\mathrm{det}$ is the normal-state detector conductance, and $e$ is the elementary charge. $f_\mathrm{L}$ and $f_\mathrm{T3}$ do not contribute since we use spin-degenerate junctions.

Figure \ref{fig:sampleModel}(d) qualitatively shows the different contributions to the nonlocal conductance. Without supercurrent, the signal comes mainly from the injected $f_\mathrm{T}$ mode (and a small $f_\mathrm{L3}$ contribution). The injected contribution is even in bias, and decays quickly as a function of contact distance and increasing magnetic field. $j_\mathrm{E}$ and $j_\mathrm{Es}$ generate contributions which are odd in bias. These contributions have the same sign in the lower Zeeman band, but opposite signs at higher energy. In the remainder of the paper, we will refer to the even and odd contributions as the ``injected'' and ``generated'' signals.

\section{Results}

\begin{figure}
    \includegraphics[width=\columnwidth]{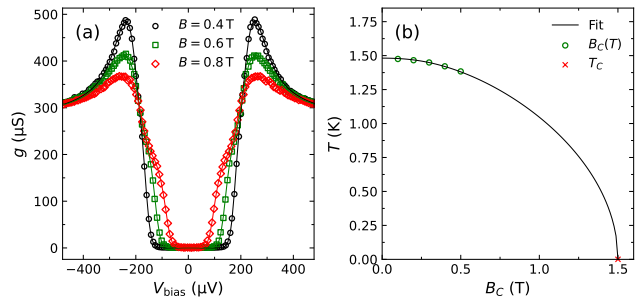}
    \caption{Sample characterization: (a) Differential conductance $g$ of one of the tunnel contacts as a function of bias voltage $V_\mathrm{bias}$ for different in-plane magnetic fields $B$. (b) Phase diagram of the superconducting wire as a function of magnetic field $B$ and temperature $T$. Symbols are obtained by measuring the resistive transition, the line is a fit explained in the text.}
    \label{fig:characterization}
\end{figure}

The sample parameters were obtained from the characterization measurements shown in Fig. \ref{fig:characterization}. Figure \ref{fig:characterization}(a) shows the differential conductance of the injector contact. The spectra were fitted to the standard model of the tunnel conductance, with an additional series resistance to account for the three-probe measurement.
The resistance of the superconducting wire was measured in a four-probe geometry using the split ends to obtain the residual resistance $R_\mathrm{4K}$ at liquid Helium temperature. The critical temperature $T_\mathrm{c}$ and critical field $B_\mathrm{c}(T)$ shown in Fig. \ref{fig:characterization}(b) were then measured by sweeping the temperature or magnetic field and taking the mid-point of the resistance of the superconducting transition. The pair potential was calculated using the BCS relation  $\Delta_0=1.74k_\mathrm{B}T_\mathrm{c}$, where $k_\mathrm{B}$ is the Boltzmann constant. The width and length of the aluminum strip as well as the detector distances were extracted from the SEM images. The nominal thickness of the aluminum wire obtained from a quartz microbalance during fabrication does not reflect the metallic cross-section relevant for the transport properties due to the unknown oxide thickness and surface roughness. Since the orbital depairing rate, and therefore the critical field, depends strongly on film thickness, we have instead determined the effective thickness by fitting the temperature dependence of the critical field (see appendix for details). The sample parameters extracted from the characterization measurements are summarized in table \ref{tab:parameters} in the appendix.

All measurement based input parameters for the numerical simulation of the nonlocal conductance measurements are gained from these characterization measurements plus the applied supercurrent.

\begin{figure}
\includegraphics[width=\columnwidth]{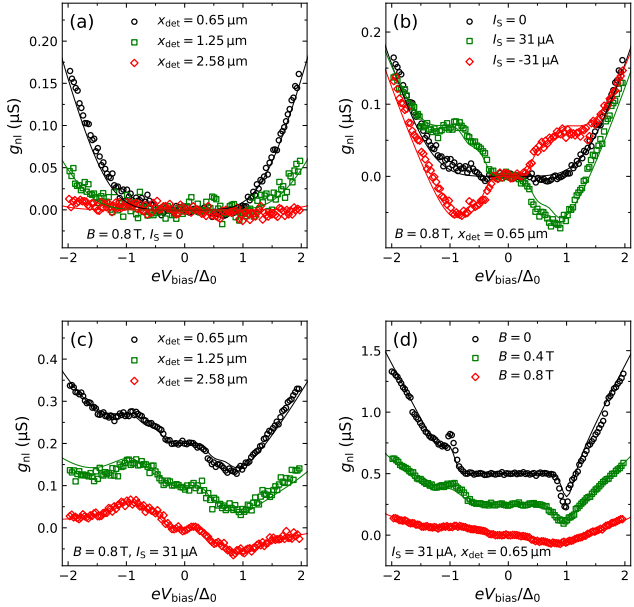}
\caption{Overview of the nonlocal differential conductance $g_\mathrm{nl}$ as a function of bias voltage. Data taken on sample A. Signals in (c) and (d) are offset vertically for better visibility. (a) $g_\mathrm{nl}$ for different detector distances without supercurrent. The signal is caused by injected charge imbalance and decays with detector distance. (b) $g_\mathrm{nl}$ for different supercurrents at fixed distance and magnetic field. (c) $g_\mathrm{nl}$ for  different detector distances. The signal from injected charge imbalance decays with distance, the supercurrent-induced part is independent of distance. (d) $g_\mathrm{nl}$ for different  magnetic fields.}
    \label{fig:overviewRaw}
\end{figure}

Figure \ref{fig:overviewRaw} is an overview of the nonlocal differential conductance $g_\mathrm{nl}$ measured on sample A (symbols) and the corresponding numerical simulations (lines). Measurements were performed for $B=0-0.8~\mathrm{T}$, below the field where the energy gap closes and up to about half of the theoretical critical current of the samples, where superconductivity starts to collapse from injection and noise of the applied current. 
Figure \ref{fig:overviewRaw}(a) shows the signal as a function of bias voltage without supercurrent for different detectors at fixed magnetic field. The signal is even in bias, and falls with increasing detector distance as the injected charge imbalance relaxes. Figure \ref{fig:overviewRaw}(b) shows the effect of supercurrent on the signal. Above the gap an additional contribution appears, which is odd in both bias and supercurrent. At higher bias, the additional signal disappears. Figure \ref{fig:overviewRaw}(c) shows the nonlocal conductance for different detectors with applied supercurrent, corresponding to the data shown in Fig.~\ref{fig:overviewRaw}(a). As in Fig.~\ref{fig:overviewRaw}(a), the even contribution from injected charge imbalance falls with detector distance while the odd contribution generated by the supercurrent is nearly independent of distance, indicating continuous creation of imbalance along the superconducting strip. Figure \ref{fig:overviewRaw}(d) shows the evolution of the signal with increasing magnetic field for a fixed detector distance. At zero field, the odd contribution consists of relatively sharp peaks at the gap. With increasing field, the signal broadens and develops a Zeeman splitting (better resolved in the upper plot of Fig.~\ref{fig:overviewRaw}(c), where the data for $B=0.8~\mathrm{T}$ are shown on a different scale). Both the even and odd contributions decrease with increasing field due to the increased charge relaxation rate.

\begin{figure}
    \includegraphics[width=\columnwidth]{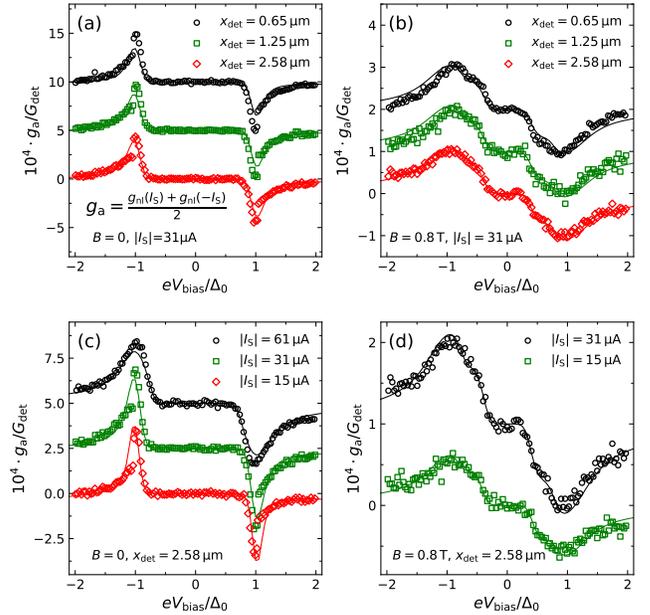}
    \caption{Overview of the antisymmetric part $g_\mathrm{a}=(g_\mathrm{nl}(I_\mathrm{S})-g_\mathrm{nl}(-I_\mathrm{S}))/2$ of the nonlocal conductance caused by quasiparticle supercurrent coupling. The signals are offset vertically for better visibility. (a) and (b) The signal at different detector distances. The signal is nearly independent of distance, indicating a continuous generation along the aluminum strip. (c) and (d) dependence of the signal on the applied supercurrent.}
    \label{fig:overviewAnalysis}
\end{figure}

To further analyze the supercurrent coupling, we use the symmetry to extract only the supercurrent-induced part of the signal, {\em i.e.}, we calculate the antisymmetric part of the conductance $g_\mathrm{a}=(g_\mathrm{nl}(I_\mathrm{S})-g_\mathrm{nl}(-I_\mathrm{S}))/2$. We also normalize the signal by $G_\mathrm{det}$ to eliminate small variations of the detector conductances. Figure \ref{fig:overviewAnalysis} is an overview over the measured and simulated $g_\mathrm{a}$.
Figure \ref{fig:overviewAnalysis}(a) and (b) show the nonlocal signal for different detector distances at $B=0$ and $B=\SI{0.8}{\tesla}$, respectively. In both cases, the signals are nearly independent of contact distance which confirms that nonequilibrium is generated continuously along the superconducting strip by supercurrent-quasiparticle coupling.
Figure \ref{fig:overviewAnalysis}(c) shows the signal at $B=0$ for three different supercurrents.
The signal is sharply peaked near the gap edge, with an increasing broadening as the supercurrent is increased. This reflects the increasing depairing by the supercurrent. Intuitively, one would expect an increase of the peak height with supercurrent, but this is compensated by the increased broadening for the differential signal shown here. The integrated signal, however, increases monotonically with supercurrent as one expects (not shown). Figure \ref{fig:overviewAnalysis}(d) shows the signal at $B=\SI{0.8}{\tesla}$ for different supercurrents. Here, the depairing due to the magnetic field is much larger than the depairing by the supercurrent, and the signal simply increases with supercurrent as expected.

\begin{figure}
    \includegraphics[width=\columnwidth]{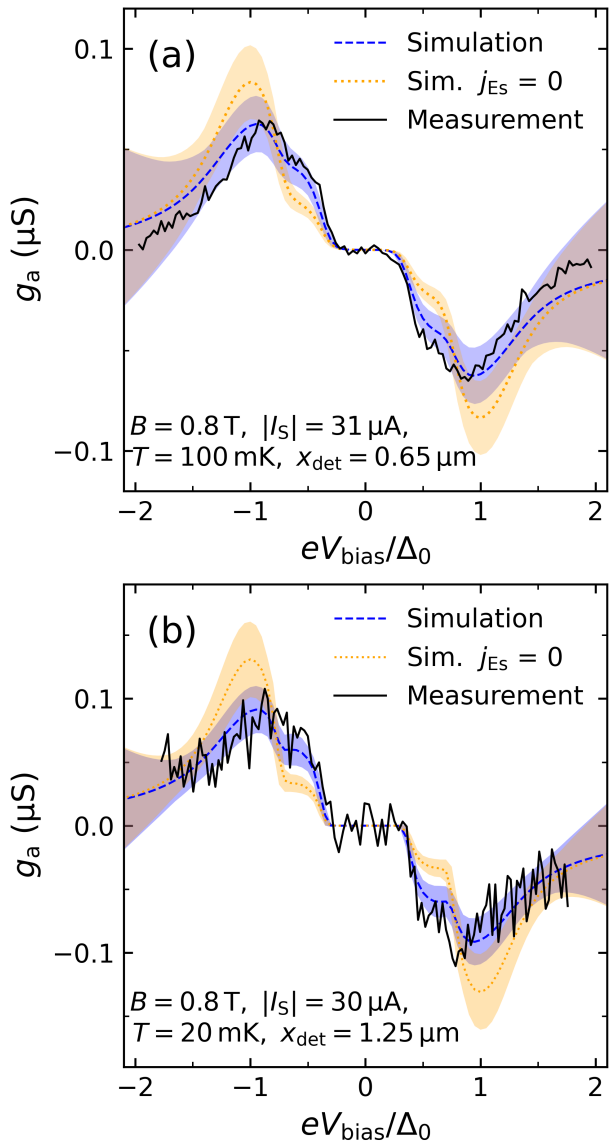}
    \caption{Comparison of $g_\mathrm{a}$ to the simulation for sample A (a) and sample  B (b). The solid lines are the measurement results, the dashed lines are the full simulations, and the dotted lines are the simulations with $j_\mathrm{Es}$ set to zero. The shaded areas are the maximum error estimate for the simulation, corresponding to the errors of the parameters.}
    \label{fig:result}
\end{figure}

Figure \ref{fig:result} shows a detailed comparison of the nonlocal signal to the numerical simulations for samples A and B. Note that all parameters were determined independently, with no free parameters left to fit. The shaded regions indicate the maximum errors determined by propagating the estimated uncertainty in sample geometry, resistances, supercurrent, $B_\mathrm{c}$ and $T_\mathrm{c}$ to the simulation result. The measured signal (solid lines) agrees with the simulation (dashed lines) within the error bars. 
The slight shift of the signal to lower bias voltage compared to the model can be explained by the reduction of the energy gap by quasiparticle injection, which is not included in the simulation. To test the effect of $j_\mathrm{Es}$, we have repeated the simulation with $j_\mathrm{Es}$ set to zero (dotted lines). These simulations do not match the data, with a downward deviation in the lower Zeeman band, and an upward deviation in the upper Zeeman band, as expected from the schematic view of signal contributions in Fig.~\ref{fig:sampleModel}(d).

\section{Discussion}

First, we would like to discuss the results in zero field, {\em i.e.}, $j_\mathrm{Es}=0$ without Zeeman splitting. The supercurrent coupling described by the term $j_\mathrm{E}\nabla\phi$ has been predicted \cite{schmid1975,schmid1979,pethick1979b,clarke1980} and experimentally confirmed \cite{clarke1979,fjordboge1981,heidel1981} in the 1970s and 80s. The historic experiments were made on cm-sized structures, much larger than the inelastic relaxation length. In this case, the $f_\mathrm{L}$ mode is given by a local equilibrium distribution with a local temperature $T(x)$, and $\nabla f_\mathrm{L}\propto \nabla T$ was created by heating one end of the wire. Also, the historic experiments were focused mainly on the temperature range close to the critical temperature, where the charge relaxation time diverges.

In contrast, the present experiments were performed on $\mathrm{\mu m}$-sized structures with tunnel injection at low temperature, where the Fermi distribution has a relatively sharp edge and inelastic scattering can be mostly neglected. As a consequence, our experiments have spectral resolution, and provide an indirect measurement of the spectral supercurrent $j_\mathrm{E}$. $j_\mathrm{E}$ is not easily accessible to experiments. The spectral supercurrent in SNS Josephson junctions has been probed by controlling the distribution function \cite{baselmans1999}, but we are not aware of similar experiments in bulk superconducting wires. For weak depairing, $j_\mathrm{E}$ is sharply peaked above the gap, and quickly drops to zero at higher energy. This behavior is reflected in the signal shown in Fig.~\ref{fig:overviewAnalysis}(c).

The generated signals are nearly independent of contact distance, as expected for a constant gradient of $f_\mathrm{L}$. The constant gradient of $f_\mathrm{L}$ in the model is the result of neglecting inelastic scattering. Inelastic scattering will eventually lead to a relaxation of $f_\mathrm{L}$, with a relaxation length of $\lambda_\mathrm{L}\approx5-10~\mathrm{\mu m}$ found in our previous experiments on similar structures \cite{huebler2012b,wolf2013,heidrich2019}. In the present experiments, the contact distances were $x_\mathrm{det}\lesssim\lambda_\mathrm{L}$, so that neglecting inelastic relaxation is justified. Also, in previous comparisons of our experiments to the model the signals could be adequately described neglecting inelastic scattering \cite{silaev2015,heidrich2019}.

At high fields, the nonlocal signals broaden due to depairing, and a double-step structure is visible due to the Zeeman splitting. Since the contributions generated by both $j_\mathrm{E}$ and $j_\mathrm{Es}$ are odd functions of supercurrent and bias, they can not by distinguished by symmetry. Instead, we have compared the signals to simulations including and excluding $j_\mathrm{Es}$, and found that neglecting $j_\mathrm{Es}$ does not describe the signal within the error bars. In particular, the relative signal weight in the upper and lower Zeeman band requires inclusion of $j_\mathrm{Es}$. This follows from the opposite sign of the contribution of $j_\mathrm{Es}$ in the upper and lower Zeeman band as shown schematically in Fig.~\ref{fig:sampleModel}(d), and is independent of any small errors in overall signal magnitude due to inaccuracies of model parameters.

To conclude, we have experimentally investigated supercurrent-induced coupling of nonequilibrium modes in high-field superconductors, and found evidence for the recently predicted spin-dependent coupling term $j_\mathrm{Es}\nabla\phi$ \cite{aikebaier2018}. The interplay of spin-dependent supercurrents and quasiparticles may find applications in superconducting spintronics.

\section*{Appendix}
The samples are modeled using the quasiclassical model for dirty superconductors, with two additional approximations. First, spectral properties are calculated for a homogeneous wire in equilibrium, and only the kinetic equations contain gradients and nonequilibrium distributions. Second, inelastic scattering is neglected, as explained in the main text. 

In the following, all energies are in units of the pair potential $\Delta_0$ at zero temperature and zero field, and lengths are measured in units of the dirty-limit coherence length $\xi=\sqrt{\hbar D_\mathrm{N}/\Delta_0}$, where $D_\mathrm{N}$ is the diffusion constant in the normal state. The spin index is $\sigma=\pm 1$, and $\sigma=+1$ corresponds to spin down $(\downarrow)$, {\em i.e.}, magnetic moment parallel to the applied magnetic field.

\paragraph{Spectral properties.} The model for the spectral properties used is based on \cite{maki1964}. The Usadel equation for a homogeneous superconductor has the form
\begin{equation}
\Delta G_\sigma + i(\varepsilon+\sigma\varepsilon_\mathrm{z})F_\sigma + \Sigma_\zeta + \Sigma_\mathrm{so} = 0,
\label{eqn:usadel}
\end{equation}
where $G_\sigma$ and $F_\sigma$ are the normal and anomalous Green's functions for spin  $\sigma$, $\varepsilon$ is the normalized energy, and $\varepsilon_\mathrm{z}$ is the normalized spin splitting. $\Delta$ is the normalized pair potential, which has to be determined self-consistently, as explained below. The Green's functions are normalized by $F_\sigma^2+G_\sigma^2=1$, which we satisfy using the parametrization $F_\sigma=\sin\left(\theta_\sigma\right)$ and $G_\sigma=\cos\left(\theta_\sigma\right)$ with the complex pairing angle $\theta_\sigma$.

The self-energy due to orbital pair breaking is given by
\begin{equation}
\Sigma_\zeta = -\zeta F_\sigma G_\sigma,
\end{equation}
where the pair-breaking parameter $\zeta$ has two contributions,
\begin{equation}
\zeta=\frac{1}{2}\left(\frac{B}{B_\mathrm{c,orb}}\right)^2+\frac{1}{2}\left(\nabla\phi\right)^2.
\end{equation}
The first term is due to the applied in-plane magnetic field \cite{maki1964}, and the second term is due to the phase gradient $\nabla\phi$ induced by the supercurrent \cite{anthore2003}.
The effect of the magnetic field is conveniently parametrized by the ``orbital'' critical field $B_\mathrm{c,orb}$, which is related to sample parameters by \cite{maki1969}
\begin{equation}
\frac{D_\mathrm{N} e^2 B_\mathrm{c,orb}^2 t^2}{6 \hbar\Delta_0}y(\frac{\pi l}{d}) = \frac{1}{2},
\label{eqn:Bcorb}
\end{equation}
where $t$ is the film thickness, $l$ is the mean free path, and
\begin{equation}
y(z)=\frac{3}{2}\frac{(1+z^2)\mathrm{arctan}(z)-z}{z^3}
\end{equation}
is a correction due to nonlocal electrodynamics. Note that the actual critical field $B_\mathrm{c}$ is smaller than $B_\mathrm{c,orb}$ due to additional depairing by the Zeeman splitting.

The self-energy due to spin-orbit scattering is
\begin{equation}
\Sigma_\mathrm{so}=-\sigma b_\mathrm{so}(F_\uparrow G_\downarrow-F_\downarrow G_\uparrow),
\end{equation}
where the spin-orbit scattering parameter is
\begin{equation}
b_\mathrm{so}=\frac{\hbar}{3\tau_\mathrm{so}\Delta},
\end{equation}
and  $\tau_\mathrm{so}$ is the spin-orbit scattering time. $b_\mathrm{so}$ can not be determined accurately from our tunnel conductance measurement. We have therefore assumed $b_\mathrm{so}=0.02$, similar to the values obtained from earlier nonlocal spin-valve experiments \cite{huebler2012b,wolf2014c} on our aluminum films.

The model is completed by the self-consistency equation for the pair potential $\Delta$ and the Zeeman splitting $\varepsilon_\mathrm{z}$ including Fermi-liquid renormalization \cite{alexander1985},
\begin{equation}
\mathrm{ln}\left(\frac{T}{T_\mathrm{c}}\right)=\frac{\omega}{\Delta}\sum_{\omega_n}\left(F_\mathrm{s}(i\omega_n)-\frac{\Delta}{\omega_n}\right),
\label{sc1}
\end{equation}
\begin{equation}
\varepsilon_\mathrm{z} - (1-A^a_0)\frac{\mu_\mathrm{B}B}{\Delta_0}=A^a_0\omega_1\sum_{\omega_n}i G_\mathrm{t}(i\omega_n),
\label{sc2}
\end{equation}
where
\begin{equation}
A^a_0=\frac{G_0}{G_0+1}.
\end{equation}
$G_0$ is the Fermi liquid parameter, and $\omega_n=(2n-1)\pi k_\mathrm{B} T/\Delta_0$ is the $n$-the Matsubara frequency. $F_\mathrm{s}=(F_\downarrow+F_\uparrow)/2$ and $G_\mathrm{t}=(G_\downarrow-G_\uparrow)/2$ are the singlet anomalous and triplet normal Green's function, respectively. Literature values for $G_0$ range from 0.16 to 0.3 \cite{catelani2008,alexander1985}, and we have assumed $G_0=0.2$. The phase transition to the normal state is always second order in our samples due to the effect of orbital depairing, and for the fits of the critical field, we have used the usual approximation of the self-consistency equations for $\Delta\rightarrow 0$ (Eq.~(85) of Ref. \onlinecite{alexander1985}).

\begin{table}[t]
    \centering
    \begin{tabular}{c|cccccccc}
    Sample & $T_\mathrm{c}$ & $B_\mathrm{c,orb}$ & $\Delta_0$ & $I_0$ & $\xi$ & $G_\mathrm{inj}$ & $G_\mathrm{det}$ \\
           & (K)            & (T)            & ($\mathrm{\mu eV}$) & ($\mathrm{\mu A}$) & (nm) & ($\mathrm{\mu S}$) & ($\mathrm{\mu S}$) \\ \hline
    A &  1.50           & 1.66            & 228                 & 121  & 122  &294 & 559 - 599 \\
    B &  1.50           & 1.63            & 228                 & 185  & 120  &766 & 731 - 742 
    \end{tabular}
    \caption{Overview of sample parameters. Critical temperature $T_\mathrm{c}$, orbital critical field $B_\mathrm{c,orb}$, pair potential $\Delta_0$, characteristic current  $I_0$, coherence length $\xi$, injector conductance $G_\mathrm{inj}$ and detector conductance $G_\mathrm{det}$.}
    \label{tab:parameters}
\end{table}

\paragraph{Kinetic equations.} Quasiparticle transport is described by four distribution functions $f_\mathrm{L}$, $f_\mathrm{T3}$, $f_\mathrm{T}$ and $f_\mathrm{L3}$, corresponding to energy, spin, charge and spin-energy currents $j_\mathrm{e}$, $j_\mathrm{s}$, $j_\mathrm{c}$ and $j_\mathrm{se}$, respectively. In equilibrium, only $f_\mathrm{L}$ is nonzero and given by
\begin{equation}
    f_0(\varepsilon)=\tanh{\left(\frac{\varepsilon}{2t}\right)},
\end{equation}
where $t=k_\mathrm{B}T/\Delta_0$ is the normalized temperature.
In the following, we will only consider the deviation from equilibrium, {\em i.e.}, we implicitly subtract $f_0$ from $f_\mathrm{L}$. 

The distribution functions and currents are related by \cite{silaev2015,bobkova2016,aikebaier2018}
\begin{equation}
\begin{pmatrix}
j_\mathrm{e}\\j_\mathrm{s}\\j_\mathrm{c}\\j_\mathrm{se}
\end{pmatrix}
=
\begin{pmatrix}
D_\mathrm{L}\nabla & D_\mathrm{T3}\nabla & j_\mathrm{E}\nabla\phi & j_\mathrm{Es}\nabla\phi \\
D_\mathrm{T3}\nabla & D_\mathrm{L}\nabla & j_\mathrm{Es}\nabla\phi & j_\mathrm{E}\nabla\phi \\
j_\mathrm{E}\nabla\phi & j_\mathrm{Es}\nabla\phi & D_\mathrm{T}\nabla & D_\mathrm{L3}\nabla \\
j_\mathrm{Es}\nabla\phi & j_\mathrm{E}\nabla\phi & D_\mathrm{L3}\nabla & D_\mathrm{T}\nabla
\end{pmatrix}
\begin{pmatrix}
f_\mathrm{L} \\ f_\mathrm{T3} \\ f_\mathrm{T} \\ f_\mathrm{L3}
\end{pmatrix}.
\label{eqn:transport1}
\end{equation}
Here, the $D_m$ are the spectral diffusion coefficients for mode $m$. The spectral supercurrent densities are
\begin{eqnarray}
    j_\mathrm{E} & = & \frac{1}{2}\mathrm{Im}\left(F_\downarrow^2+F_\uparrow^2\right), \\
    j_\mathrm{Es} & = & \frac{1}{2}\mathrm{Im}\left(F_\downarrow^2-F_\uparrow^2\right).
\end{eqnarray}
Current densities are either driven by gradients of the distribution functions, or through their coupling to the supercurrent. Measuring the off-diagonal coupling due to $j_\mathrm{Es}$ is the goal of this paper.

Relaxation of the nonequilibrium currents is given by
\begin{equation}
\begin{split}
\nabla j_\mathrm{e} & = 0\\
\nabla j_\mathrm{s} & = S_\mathrm{T3}f_\mathrm{T3}\\
\nabla j_\mathrm{c} & = R_\mathrm{T} f_\mathrm{T} + R_\mathrm{L3}f_\mathrm{L3}\\
\nabla j_\mathrm{se} & = (R_\mathrm{T}+S_\mathrm{L3})f_\mathrm{L3} +  R_\mathrm{L3}f_\mathrm{T}
\end{split}
\label{transport2}
\end{equation}
$S_\mathrm{T3}$ and $S_\mathrm{L3}$ are spin relaxation rates due to spin-orbit scattering (we neglect spin flips by magnetic impurities). The coefficients $R_\mathrm{T}$ and $R_\mathrm{L3}$ describe charge relaxation by coupling to the superconducting condensate. As described above, we neglect inelastic scattering, and therefore the relaxation rate of $j_\mathrm{e}$ is zero.

The differential equations are supplemented by boundary conditions. For a spin-degenerate injector at $x=0$, these read
\begin{equation}
\begin{pmatrix}
\left[j_\mathrm{e}\right]\\\left[j_\mathrm{s}\right]\\\left[j_\mathrm{c}\right]\\\left[j_\mathrm{se}\right]
\end{pmatrix}
= \kappa_\mathrm{I}
  \begin{pmatrix}
    N_+ & N_- & 0 & 0 \\
    N_- & N_+ & 0 & 0 \\
    0 & 0 & N_+ & N_- \\
    0 & 0 & N_- & N_+
  \end{pmatrix}
\begin{pmatrix}
\left[f_\mathrm{L}\right] \\ \left[f_\mathrm{T3}\right] \\ \left[f_\mathrm{T}\right] \\ \left[f_\mathrm{L3}\right]
\end{pmatrix},
\label{eqn:bc}
\end{equation}
where $\left[j_m\right]=j_m(x=0+)-j_m(x=0-)$ and $\left[f_m\right]=f_m(x=0)-f_m^{\mathrm{inj}}$. The effective injection rate is given by
\begin{equation}
    \kappa_\mathrm{I}=G_{\mathrm{inj}}\frac{\rho_\mathrm{N}\xi}{A},
\end{equation}
where $A$ is the cross-section of the wire, and $\rho_\mathrm{N}$ is the normal-state resistivity.
The distribution functions of the injector are given by
\begin{eqnarray}
f_{L}^{\mathrm{inj}} & = & \frac{1}{2}\left(f_0\left(\varepsilon+\mu\right)+f_0\left(\varepsilon-\mu\right)\right)-f_0\left(\varepsilon\right), \\
f_{T}^{\mathrm{inj}} & = & \frac{1}{2}\left(f_0\left(\varepsilon+\mu\right)-f_0\left(\varepsilon-\mu\right)\right),
\end{eqnarray}
where $\mu=-eV_\mathrm{inj}/\Delta_0$ is the electrochemical potential of the injector. The T3 and L3 modes are zero in the injector.
The boundary conditions at the ends of the wire are $f_m(x=\pm L)=0$. 

\begin{figure}
    \includegraphics[width=0.7\columnwidth]{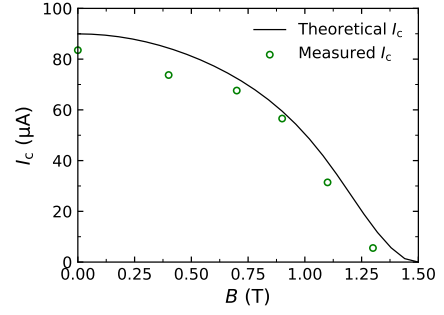}
    \caption{Comparison of the measured (symbols) and theoretical (line) critical current of sample A at $T=100~\mathrm{mK}$.}
    \label{fig:Ic}
\end{figure}

\paragraph{Observables.}
The spin-resolved density of states is $N_\sigma=\mathrm{Re}(G_\sigma)$, which gives the spin-symmetric and antisymmetric parts
\begin{equation}
N_\pm = N_\downarrow\pm N_\uparrow.
\end{equation}
The differential tunnel conductance of the injector is
\begin{equation}
g = \frac{G_\mathrm{inj}}{2}\int_{-\infty}^{\infty}N_+\frac{\partial f_0(\epsilon-\mu)}{\partial \mu}d\epsilon.
\end{equation}
The supercurrent is  given by
\begin{equation}
I_\mathrm{S} = I_0\frac{\nabla\phi}{2}\int_{-\infty}^\infty f_0\left(\varepsilon\right)j_\mathrm{E}  d\varepsilon
\end{equation}
with the characteristic current
\begin{equation} 
I_0  = \frac{\Delta_0 A}{e \rho_\mathrm{N} \xi}.
\end{equation}
The theoretical critical current at $T=0$ and $B=0$ is about $0.75I_\mathrm{0}$.
Fig.~\ref{fig:Ic} compares the measured critical current for sample A to the model prediction. The measured critical current is slightly smaller than predicted, probably as a result of premature escape due to noise.

For the simulations, first the spectral properties were calculated self-consistently for the applied field and temperature, including $\nabla\phi$ determined self-consistently from the applied supercurrent. Then the kinetic equations were solved numerically on an equidistant position and energy grid with spacings $\delta x = L/80$ and $\delta \varepsilon=1/20$, respectively.

\bibliography{sdsc}

\begin{thebibliography}{44}%
\makeatletter
\providecommand \@ifxundefined [1]{%
 \@ifx{#1\undefined}
}%
\providecommand \@ifnum [1]{%
 \ifnum #1\expandafter \@firstoftwo
 \else \expandafter \@secondoftwo
 \fi
}%
\providecommand \@ifx [1]{%
 \ifx #1\expandafter \@firstoftwo
 \else \expandafter \@secondoftwo
 \fi
}%
\providecommand \natexlab [1]{#1}%
\providecommand \enquote  [1]{``#1''}%
\providecommand \bibnamefont  [1]{#1}%
\providecommand \bibfnamefont [1]{#1}%
\providecommand \citenamefont [1]{#1}%
\providecommand \href@noop [0]{\@secondoftwo}%
\providecommand \href [0]{\begingroup \@sanitize@url \@href}%
\providecommand \@href[1]{\@@startlink{#1}\@@href}%
\providecommand \@@href[1]{\endgroup#1\@@endlink}%
\providecommand \@sanitize@url [0]{\catcode `\\12\catcode `\$12\catcode
  `\&12\catcode `\#12\catcode `\^12\catcode `\_12\catcode `\%12\relax}%
\providecommand \@@startlink[1]{}%
\providecommand \@@endlink[0]{}%
\providecommand \url  [0]{\begingroup\@sanitize@url \@url }%
\providecommand \@url [1]{\endgroup\@href {#1}{\urlprefix }}%
\providecommand \urlprefix  [0]{URL }%
\providecommand \Eprint [0]{\href }%
\providecommand \doibase [0]{https://doi.org/}%
\providecommand \selectlanguage [0]{\@gobble}%
\providecommand \bibinfo  [0]{\@secondoftwo}%
\providecommand \bibfield  [0]{\@secondoftwo}%
\providecommand \translation [1]{[#1]}%
\providecommand \BibitemOpen [0]{}%
\providecommand \bibitemStop [0]{}%
\providecommand \bibitemNoStop [0]{.\EOS\space}%
\providecommand \EOS [0]{\spacefactor3000\relax}%
\providecommand \BibitemShut  [1]{\csname bibitem#1\endcsname}%
\let\auto@bib@innerbib\@empty
\bibitem [{\citenamefont {Langenberg}\ and\ \citenamefont
  {Larkin}(1986)}]{langenberg1986}%
  \BibitemOpen
  \bibinfo {editor} {\bibfnamefont {D.~N.}\ \bibnamefont {Langenberg}}\ and\
  \bibinfo {editor} {\bibfnamefont {A.~I.}\ \bibnamefont {Larkin}},\ eds.,\
  \href@noop {} {\emph {\bibinfo {title} {Nonequilibrium
  {{Superconductivity}}}}}\ (\bibinfo  {publisher} {{North Holland}},\ \bibinfo
  {address} {{Amsterdam}},\ \bibinfo {year} {1986})\BibitemShut {NoStop}%
\bibitem [{\citenamefont {Schmid}\ and\ \citenamefont
  {Sch{\"o}n}(1975)}]{schmid1975}%
  \BibitemOpen
  \bibfield  {author} {\bibinfo {author} {\bibfnamefont {A.}~\bibnamefont
  {Schmid}}\ and\ \bibinfo {author} {\bibfnamefont {G.}~\bibnamefont
  {Sch{\"o}n}},\ }\bibfield  {title} {\bibinfo {title} {Linearized kinetic
  equations and relaxation processes of a superconductor near {{$T_c$}}},\
  }\href {https://doi.org/10.1007/BF00115264} {\bibfield  {journal} {\bibinfo
  {journal} {J. Low Temp. Phys.}\ }\textbf {\bibinfo {volume} {20}},\ \bibinfo
  {pages} {207} (\bibinfo {year} {1975})}\BibitemShut {NoStop}%
\bibitem [{\citenamefont {Schmid}\ and\ \citenamefont
  {Sch{\"o}n}(1979)}]{schmid1979}%
  \BibitemOpen
  \bibfield  {author} {\bibinfo {author} {\bibfnamefont {A.}~\bibnamefont
  {Schmid}}\ and\ \bibinfo {author} {\bibfnamefont {G.}~\bibnamefont
  {Sch{\"o}n}},\ }\bibfield  {title} {\bibinfo {title} {Generation of {{Branch
  Imbalance}} by the {{Interaction}} between {{Supercurrent}} and {{Thermal
  Gradient}}},\ }\href {https://doi.org/10.1103/PhysRevLett.43.793} {\bibfield
  {journal} {\bibinfo  {journal} {Phys. Rev. Lett.}\ }\textbf {\bibinfo
  {volume} {43}},\ \bibinfo {pages} {793} (\bibinfo {year} {1979})}\BibitemShut
  {NoStop}%
\bibitem [{\citenamefont {Pethick}\ and\ \citenamefont
  {Smith}(1979)}]{pethick1979b}%
  \BibitemOpen
  \bibfield  {author} {\bibinfo {author} {\bibfnamefont {C.~J.}\ \bibnamefont
  {Pethick}}\ and\ \bibinfo {author} {\bibfnamefont {H.}~\bibnamefont
  {Smith}},\ }\bibfield  {title} {\bibinfo {title} {Generation of {{Charge
  Imbalance}} in a {{Superconductor}} by a {{Temperature Gradient}}},\ }\href
  {https://doi.org/10.1103/PhysRevLett.43.640} {\bibfield  {journal} {\bibinfo
  {journal} {Phys. Rev. Lett.}\ }\textbf {\bibinfo {volume} {43}},\ \bibinfo
  {pages} {640} (\bibinfo {year} {1979})}\BibitemShut {NoStop}%
\bibitem [{\citenamefont {Clarke}\ and\ \citenamefont
  {Tinkham}(1980)}]{clarke1980}%
  \BibitemOpen
  \bibfield  {author} {\bibinfo {author} {\bibfnamefont {J.}~\bibnamefont
  {Clarke}}\ and\ \bibinfo {author} {\bibfnamefont {M.}~\bibnamefont
  {Tinkham}},\ }\bibfield  {title} {\bibinfo {title} {Theory of {{Quasiparticle
  Charge Imbalance Induced}} in a {{Superconductor}} by a {{Supercurrent}} in
  the {{Presence}} of a {{Thermal Gradient}}},\ }\href
  {https://doi.org/10.1103/PhysRevLett.44.106} {\bibfield  {journal} {\bibinfo
  {journal} {Phys. Rev. Lett.}\ }\textbf {\bibinfo {volume} {44}},\ \bibinfo
  {pages} {106} (\bibinfo {year} {1980})}\BibitemShut {NoStop}%
\bibitem [{\citenamefont {Clarke}\ \emph {et~al.}(1979)\citenamefont {Clarke},
  \citenamefont {Fjordb{\o}ge},\ and\ \citenamefont {Lindelof}}]{clarke1979}%
  \BibitemOpen
  \bibfield  {author} {\bibinfo {author} {\bibfnamefont {J.}~\bibnamefont
  {Clarke}}, \bibinfo {author} {\bibfnamefont {B.~R.}\ \bibnamefont
  {Fjordb{\o}ge}},\ and\ \bibinfo {author} {\bibfnamefont {P.~E.}\ \bibnamefont
  {Lindelof}},\ }\bibfield  {title} {\bibinfo {title} {Supercurrent-induced
  charge imbalance measured in a superconductor in the presence of a thermal
  gradient},\ }\href {https://doi.org/10.1103/PhysRevLett.43.642} {\bibfield
  {journal} {\bibinfo  {journal} {Phys. Rev. Lett.}\ }\textbf {\bibinfo
  {volume} {43}},\ \bibinfo {pages} {642} (\bibinfo {year} {1979})}\BibitemShut
  {NoStop}%
\bibitem [{\citenamefont {Fjordb{\o}ge}\ \emph {et~al.}(1981)\citenamefont
  {Fjordb{\o}ge}, \citenamefont {Lindelof},\ and\ \citenamefont
  {Clarke}}]{fjordboge1981}%
  \BibitemOpen
  \bibfield  {author} {\bibinfo {author} {\bibfnamefont {B.~R.}\ \bibnamefont
  {Fjordb{\o}ge}}, \bibinfo {author} {\bibfnamefont {P.~E.}\ \bibnamefont
  {Lindelof}},\ and\ \bibinfo {author} {\bibfnamefont {J.}~\bibnamefont
  {Clarke}},\ }\bibfield  {title} {\bibinfo {title} {Charge imbalance in
  superconducting tin films produced by a supercurrent in the presence of a
  temperature gradient},\ }\href {https://doi.org/10.1007/BF00117842}
  {\bibfield  {journal} {\bibinfo  {journal} {J. Low Temp. Phys.}\ }\textbf
  {\bibinfo {volume} {44}},\ \bibinfo {pages} {535} (\bibinfo {year}
  {1981})}\BibitemShut {NoStop}%
\bibitem [{\citenamefont {Heidel}\ and\ \citenamefont
  {Garland}(1981)}]{heidel1981}%
  \BibitemOpen
  \bibfield  {author} {\bibinfo {author} {\bibfnamefont {D.~F.}\ \bibnamefont
  {Heidel}}\ and\ \bibinfo {author} {\bibfnamefont {J.~C.}\ \bibnamefont
  {Garland}},\ }\bibfield  {title} {\bibinfo {title} {Thermoelectric charge
  imbalance in superconducting aluminum},\ }\href
  {https://doi.org/10.1007/BF00120779} {\bibfield  {journal} {\bibinfo
  {journal} {J. Low Temp. Phys.}\ }\textbf {\bibinfo {volume} {44}},\ \bibinfo
  {pages} {295} (\bibinfo {year} {1981})}\BibitemShut {NoStop}%
\bibitem [{\citenamefont {Johnson}(1994)}]{johnson1994}%
  \BibitemOpen
  \bibfield  {author} {\bibinfo {author} {\bibfnamefont {M.}~\bibnamefont
  {Johnson}},\ }\bibfield  {title} {\bibinfo {title} {Spin coupled resistance
  observed in ferromagnet-superconductor-ferromagnet trilayers},\ }\href
  {https://doi.org/10.1063/1.112015} {\bibfield  {journal} {\bibinfo  {journal}
  {Appl. Phys. Lett.}\ }\textbf {\bibinfo {volume} {65}},\ \bibinfo {pages}
  {1460} (\bibinfo {year} {1994})}\BibitemShut {NoStop}%
\bibitem [{\citenamefont {Poli}\ \emph {et~al.}(2008)\citenamefont {Poli},
  \citenamefont {Morten}, \citenamefont {Urech}, \citenamefont {Brataas},
  \citenamefont {Haviland},\ and\ \citenamefont {Korenivski}}]{poli2008}%
  \BibitemOpen
  \bibfield  {author} {\bibinfo {author} {\bibfnamefont {N.}~\bibnamefont
  {Poli}}, \bibinfo {author} {\bibfnamefont {J.~P.}\ \bibnamefont {Morten}},
  \bibinfo {author} {\bibfnamefont {M.}~\bibnamefont {Urech}}, \bibinfo
  {author} {\bibfnamefont {A.}~\bibnamefont {Brataas}}, \bibinfo {author}
  {\bibfnamefont {D.~B.}\ \bibnamefont {Haviland}},\ and\ \bibinfo {author}
  {\bibfnamefont {V.}~\bibnamefont {Korenivski}},\ }\bibfield  {title}
  {\bibinfo {title} {Spin {{Injection}} and {{Relaxation}} in a {{Mesoscopic
  Superconductor}}},\ }\href {https://doi.org/10.1103/PhysRevLett.100.136601}
  {\bibfield  {journal} {\bibinfo  {journal} {Phys. Rev. Lett.}\ }\textbf
  {\bibinfo {volume} {100}},\ \bibinfo {pages} {136601} (\bibinfo {year}
  {2008})}\BibitemShut {NoStop}%
\bibitem [{\citenamefont {Yang}\ \emph {et~al.}(2010)\citenamefont {Yang},
  \citenamefont {Yang}, \citenamefont {Takahashi}, \citenamefont {Maekawa},\
  and\ \citenamefont {Parkin}}]{yang2010}%
  \BibitemOpen
  \bibfield  {author} {\bibinfo {author} {\bibfnamefont {H.}~\bibnamefont
  {Yang}}, \bibinfo {author} {\bibfnamefont {S.-H.}\ \bibnamefont {Yang}},
  \bibinfo {author} {\bibfnamefont {S.}~\bibnamefont {Takahashi}}, \bibinfo
  {author} {\bibfnamefont {S.}~\bibnamefont {Maekawa}},\ and\ \bibinfo {author}
  {\bibfnamefont {S.~S.~P.}\ \bibnamefont {Parkin}},\ }\bibfield  {title}
  {\bibinfo {title} {Extremely long quasiparticle spin lifetimes in
  superconducting aluminium using {{MgO}} tunnel spin injectors},\ }\href
  {https://doi.org/10.1038/nmat2781} {\bibfield  {journal} {\bibinfo  {journal}
  {Nat. Mater.}\ }\textbf {\bibinfo {volume} {9}},\ \bibinfo {pages} {586}
  (\bibinfo {year} {2010})}\BibitemShut {NoStop}%
\bibitem [{\citenamefont {Wakamura}\ \emph {et~al.}(2015)\citenamefont
  {Wakamura}, \citenamefont {Akaike}, \citenamefont {Omori}, \citenamefont
  {Niimi}, \citenamefont {Takahashi}, \citenamefont {Fujimaki}, \citenamefont
  {Maekawa},\ and\ \citenamefont {Otani}}]{wakamura2015}%
  \BibitemOpen
  \bibfield  {author} {\bibinfo {author} {\bibfnamefont {T.}~\bibnamefont
  {Wakamura}}, \bibinfo {author} {\bibfnamefont {H.}~\bibnamefont {Akaike}},
  \bibinfo {author} {\bibfnamefont {Y.}~\bibnamefont {Omori}}, \bibinfo
  {author} {\bibfnamefont {Y.}~\bibnamefont {Niimi}}, \bibinfo {author}
  {\bibfnamefont {S.}~\bibnamefont {Takahashi}}, \bibinfo {author}
  {\bibfnamefont {A.}~\bibnamefont {Fujimaki}}, \bibinfo {author}
  {\bibfnamefont {S.}~\bibnamefont {Maekawa}},\ and\ \bibinfo {author}
  {\bibfnamefont {Y.}~\bibnamefont {Otani}},\ }\bibfield  {title} {\bibinfo
  {title} {Quasiparticle-mediated spin {{Hall}} effect in a superconductor},\
  }\href {https://doi.org/10.1038/nmat4276} {\bibfield  {journal} {\bibinfo
  {journal} {Nat. Mater.}\ }\textbf {\bibinfo {volume} {14}},\ \bibinfo {pages}
  {675} (\bibinfo {year} {2015})}\BibitemShut {NoStop}%
\bibitem [{\citenamefont {Beckmann}(2016)}]{beckmann2016}%
  \BibitemOpen
  \bibfield  {author} {\bibinfo {author} {\bibfnamefont {D.}~\bibnamefont
  {Beckmann}},\ }\bibfield  {title} {\bibinfo {title} {Spin manipulation in
  nanoscale superconductors},\ }\href
  {https://doi.org/10.1088/0953-8984/28/16/163001} {\bibfield  {journal}
  {\bibinfo  {journal} {J. Phys.: Condens. Matter}\ }\textbf {\bibinfo {volume}
  {28}},\ \bibinfo {pages} {163001} (\bibinfo {year} {2016})}\BibitemShut
  {NoStop}%
\bibitem [{\citenamefont {Morten}\ \emph {et~al.}(2004)\citenamefont {Morten},
  \citenamefont {Brataas},\ and\ \citenamefont {Belzig}}]{morten2004}%
  \BibitemOpen
  \bibfield  {author} {\bibinfo {author} {\bibfnamefont {J.~P.}\ \bibnamefont
  {Morten}}, \bibinfo {author} {\bibfnamefont {A.}~\bibnamefont {Brataas}},\
  and\ \bibinfo {author} {\bibfnamefont {W.}~\bibnamefont {Belzig}},\
  }\bibfield  {title} {\bibinfo {title} {Spin transport in diffusive
  superconductors},\ }\href {https://doi.org/10.1103/PhysRevB.70.212508}
  {\bibfield  {journal} {\bibinfo  {journal} {Phys. Rev. B}\ }\textbf {\bibinfo
  {volume} {70}},\ \bibinfo {pages} {212508} (\bibinfo {year}
  {2004})}\BibitemShut {NoStop}%
\bibitem [{\citenamefont {Linder}\ and\ \citenamefont
  {Robinson}(2015)}]{linder2015}%
  \BibitemOpen
  \bibfield  {author} {\bibinfo {author} {\bibfnamefont {J.}~\bibnamefont
  {Linder}}\ and\ \bibinfo {author} {\bibfnamefont {J.~W.~A.}\ \bibnamefont
  {Robinson}},\ }\bibfield  {title} {\bibinfo {title} {Superconducting
  spintronics},\ }\href {https://doi.org/10.1038/nphys3242} {\bibfield
  {journal} {\bibinfo  {journal} {Nat. Phys.}\ }\textbf {\bibinfo {volume}
  {11}},\ \bibinfo {pages} {307} (\bibinfo {year} {2015})}\BibitemShut
  {NoStop}%
\bibitem [{\citenamefont {Eschrig}(2015)}]{eschrig2015}%
  \BibitemOpen
  \bibfield  {author} {\bibinfo {author} {\bibfnamefont {M.}~\bibnamefont
  {Eschrig}},\ }\bibfield  {title} {\bibinfo {title} {Spin-polarized
  supercurrents for spintronics: A review of current progress},\ }\href
  {https://doi.org/10.1088/0034-4885/78/10/104501} {\bibfield  {journal}
  {\bibinfo  {journal} {Rep. Prog. Phys.}\ }\textbf {\bibinfo {volume} {78}},\
  \bibinfo {pages} {104501} (\bibinfo {year} {2015})}\BibitemShut {NoStop}%
\bibitem [{\citenamefont {Bobkova}\ and\ \citenamefont
  {Bobkov}(2010)}]{bobkova2010}%
  \BibitemOpen
  \bibfield  {author} {\bibinfo {author} {\bibfnamefont {I.~V.}\ \bibnamefont
  {Bobkova}}\ and\ \bibinfo {author} {\bibfnamefont {A.~M.}\ \bibnamefont
  {Bobkov}},\ }\bibfield  {title} {\bibinfo {title} {Triplet contribution to
  the {{Josephson}} current in the nonequilibrium
  superconductor/ferromagnet/superconductor junction},\ }\href
  {https://doi.org/10.1103/PhysRevB.82.024515} {\bibfield  {journal} {\bibinfo
  {journal} {Phys. Rev. B}\ }\textbf {\bibinfo {volume} {82}},\ \bibinfo
  {pages} {024515} (\bibinfo {year} {2010})}\BibitemShut {NoStop}%
\bibitem [{\citenamefont {Bobkov}\ and\ \citenamefont
  {Bobkova}(2011)}]{bobkov2011}%
  \BibitemOpen
  \bibfield  {author} {\bibinfo {author} {\bibfnamefont {A.~M.}\ \bibnamefont
  {Bobkov}}\ and\ \bibinfo {author} {\bibfnamefont {I.~V.}\ \bibnamefont
  {Bobkova}},\ }\bibfield  {title} {\bibinfo {title} {Influence of
  spin-dependent quasiparticle distribution on the {{Josephson}} current
  through a ferromagnetic weak link},\ }\href
  {https://doi.org/10.1103/PhysRevB.84.054533} {\bibfield  {journal} {\bibinfo
  {journal} {Phys. Rev. B}\ }\textbf {\bibinfo {volume} {84}},\ \bibinfo
  {pages} {054533} (\bibinfo {year} {2011})}\BibitemShut {NoStop}%
\bibitem [{\citenamefont {Amundsen}\ and\ \citenamefont
  {Linder}(2020)}]{amundsen2020}%
  \BibitemOpen
  \bibfield  {author} {\bibinfo {author} {\bibfnamefont {M.}~\bibnamefont
  {Amundsen}}\ and\ \bibinfo {author} {\bibfnamefont {J.}~\bibnamefont
  {Linder}},\ }\bibfield  {title} {\bibinfo {title} {Spin accumulation induced
  by a singlet supercurrent},\ }\href
  {https://doi.org/10.1103/PhysRevB.102.100506} {\bibfield  {journal} {\bibinfo
   {journal} {Phys. Rev. B}\ }\textbf {\bibinfo {volume} {102}},\ \bibinfo
  {pages} {100506(R)} (\bibinfo {year} {2020})}\BibitemShut {NoStop}%
\bibitem [{\citenamefont {Meservey}\ \emph {et~al.}(1970)\citenamefont
  {Meservey}, \citenamefont {Tedrow},\ and\ \citenamefont
  {Fulde}}]{meservey1970}%
  \BibitemOpen
  \bibfield  {author} {\bibinfo {author} {\bibfnamefont {R.}~\bibnamefont
  {Meservey}}, \bibinfo {author} {\bibfnamefont {P.~M.}\ \bibnamefont
  {Tedrow}},\ and\ \bibinfo {author} {\bibfnamefont {P.}~\bibnamefont
  {Fulde}},\ }\bibfield  {title} {\bibinfo {title} {Magnetic {{Field
  Splitting}} of the {{Quasiparticle States}} in {{Superconducting Aluminum
  Films}}},\ }\href {https://doi.org/10.1103/PhysRevLett.25.1270} {\bibfield
  {journal} {\bibinfo  {journal} {Phys. Rev. Lett.}\ }\textbf {\bibinfo
  {volume} {25}},\ \bibinfo {pages} {1270} (\bibinfo {year}
  {1970})}\BibitemShut {NoStop}%
\bibitem [{\citenamefont {H{\"u}bler}\ \emph {et~al.}(2012)\citenamefont
  {H{\"u}bler}, \citenamefont {Wolf}, \citenamefont {Beckmann},\ and\
  \citenamefont {{v. L{\"o}hneysen}}}]{huebler2012b}%
  \BibitemOpen
  \bibfield  {author} {\bibinfo {author} {\bibfnamefont {F.}~\bibnamefont
  {H{\"u}bler}}, \bibinfo {author} {\bibfnamefont {M.~J.}\ \bibnamefont
  {Wolf}}, \bibinfo {author} {\bibfnamefont {D.}~\bibnamefont {Beckmann}},\
  and\ \bibinfo {author} {\bibfnamefont {H.}~\bibnamefont {{v.
  L{\"o}hneysen}}},\ }\bibfield  {title} {\bibinfo {title} {Long-{{Range
  Spin-Polarized Quasiparticle Transport}} in {{Mesoscopic Al Superconductors}}
  with a {{Zeeman Splitting}}},\ }\href
  {https://doi.org/10.1103/PhysRevLett.109.207001} {\bibfield  {journal}
  {\bibinfo  {journal} {Phys. Rev. Lett.}\ }\textbf {\bibinfo {volume} {109}},\
  \bibinfo {pages} {207001} (\bibinfo {year} {2012})}\BibitemShut {NoStop}%
\bibitem [{\citenamefont {Quay}\ \emph {et~al.}(2013)\citenamefont {Quay},
  \citenamefont {Chevallier}, \citenamefont {Bena},\ and\ \citenamefont
  {Aprili}}]{quay2013}%
  \BibitemOpen
  \bibfield  {author} {\bibinfo {author} {\bibfnamefont {C.~H.~L.}\
  \bibnamefont {Quay}}, \bibinfo {author} {\bibfnamefont {D.}~\bibnamefont
  {Chevallier}}, \bibinfo {author} {\bibfnamefont {C.}~\bibnamefont {Bena}},\
  and\ \bibinfo {author} {\bibfnamefont {M.}~\bibnamefont {Aprili}},\
  }\bibfield  {title} {\bibinfo {title} {Spin imbalance and spin-charge
  separation in a mesoscopic superconductor},\ }\href
  {https://doi.org/10.1038/nphys2518} {\bibfield  {journal} {\bibinfo
  {journal} {Nat. Phys.}\ }\textbf {\bibinfo {volume} {9}},\ \bibinfo {pages}
  {84} (\bibinfo {year} {2013})}\BibitemShut {NoStop}%
\bibitem [{\citenamefont {Silaev}\ \emph {et~al.}(2015)\citenamefont {Silaev},
  \citenamefont {Virtanen}, \citenamefont {Bergeret},\ and\ \citenamefont
  {Heikkil\"a}}]{silaev2015}%
  \BibitemOpen
  \bibfield  {author} {\bibinfo {author} {\bibfnamefont {M.}~\bibnamefont
  {Silaev}}, \bibinfo {author} {\bibfnamefont {P.}~\bibnamefont {Virtanen}},
  \bibinfo {author} {\bibfnamefont {F.~S.}\ \bibnamefont {Bergeret}},\ and\
  \bibinfo {author} {\bibfnamefont {T.~T.}\ \bibnamefont {Heikkil\"a}},\
  }\bibfield  {title} {\bibinfo {title} {Long-range spin accumulation from heat
  injection in mesoscopic superconductors with {{Zeeman}} splitting},\ }\href
  {https://doi.org/10.1103/PhysRevLett.114.167002} {\bibfield  {journal}
  {\bibinfo  {journal} {Phys. Rev. Lett.}\ }\textbf {\bibinfo {volume} {114}},\
  \bibinfo {pages} {167002} (\bibinfo {year} {2015})}\BibitemShut {NoStop}%
\bibitem [{\citenamefont {Krishtop}\ \emph {et~al.}(2015)\citenamefont
  {Krishtop}, \citenamefont {Houzet},\ and\ \citenamefont
  {Meyer}}]{krishtop2015}%
  \BibitemOpen
  \bibfield  {author} {\bibinfo {author} {\bibfnamefont {T.}~\bibnamefont
  {Krishtop}}, \bibinfo {author} {\bibfnamefont {M.}~\bibnamefont {Houzet}},\
  and\ \bibinfo {author} {\bibfnamefont {J.~S.}\ \bibnamefont {Meyer}},\
  }\bibfield  {title} {\bibinfo {title} {Nonequilibrium spin transport in
  {{Zeeman-split}} superconductors},\ }\href
  {https://doi.org/10.1103/PhysRevB.91.121407} {\bibfield  {journal} {\bibinfo
  {journal} {Phys. Rev. B}\ }\textbf {\bibinfo {volume} {91}},\ \bibinfo
  {pages} {121407(R)} (\bibinfo {year} {2015})}\BibitemShut {NoStop}%
\bibitem [{\citenamefont {Bobkova}\ and\ \citenamefont
  {Bobkov}(2015)}]{bobkova2015a}%
  \BibitemOpen
  \bibfield  {author} {\bibinfo {author} {\bibfnamefont {I.~V.}\ \bibnamefont
  {Bobkova}}\ and\ \bibinfo {author} {\bibfnamefont {A.~M.}\ \bibnamefont
  {Bobkov}},\ }\bibfield  {title} {\bibinfo {title} {Long-range spin imbalance
  in mesoscopic superconductors under {{Zeeman}} splitting},\ }\href
  {https://doi.org/10.1134/S0021364015020022} {\bibfield  {journal} {\bibinfo
  {journal} {JETP Lett.}\ }\textbf {\bibinfo {volume} {101}},\ \bibinfo {pages}
  {118} (\bibinfo {year} {2015})}\BibitemShut {NoStop}%
\bibitem [{\citenamefont {Bergeret}\ \emph {et~al.}(2018)\citenamefont
  {Bergeret}, \citenamefont {Silaev}, \citenamefont {Virtanen},\ and\
  \citenamefont {Heikkil{\"a}}}]{bergeret2018}%
  \BibitemOpen
  \bibfield  {author} {\bibinfo {author} {\bibfnamefont {F.~S.}\ \bibnamefont
  {Bergeret}}, \bibinfo {author} {\bibfnamefont {M.}~\bibnamefont {Silaev}},
  \bibinfo {author} {\bibfnamefont {P.}~\bibnamefont {Virtanen}},\ and\
  \bibinfo {author} {\bibfnamefont {T.~T.}\ \bibnamefont {Heikkil{\"a}}},\
  }\bibfield  {title} {\bibinfo {title} {Colloquium: {{Nonequilibrium}} effects
  in superconductors with a spin-splitting field},\ }\href
  {https://doi.org/10.1103/RevModPhys.90.041001} {\bibfield  {journal}
  {\bibinfo  {journal} {Rev. Mod. Phys.}\ }\textbf {\bibinfo {volume} {90}},\
  \bibinfo {pages} {041001} (\bibinfo {year} {2018})}\BibitemShut {NoStop}%
\bibitem [{\citenamefont {Heikkil{\"a}}\ \emph {et~al.}(2019)\citenamefont
  {Heikkil{\"a}}, \citenamefont {Silaev}, \citenamefont {Virtanen},\ and\
  \citenamefont {Bergeret}}]{heikkila2019}%
  \BibitemOpen
  \bibfield  {author} {\bibinfo {author} {\bibfnamefont {T.~T.}\ \bibnamefont
  {Heikkil{\"a}}}, \bibinfo {author} {\bibfnamefont {M.}~\bibnamefont
  {Silaev}}, \bibinfo {author} {\bibfnamefont {P.}~\bibnamefont {Virtanen}},\
  and\ \bibinfo {author} {\bibfnamefont {F.~S.}\ \bibnamefont {Bergeret}},\
  }\bibfield  {title} {\bibinfo {title} {Thermal, electric and spin transport
  in superconductor/ferromagnetic-insulator structures},\ }\href
  {https://doi.org/10.1016/j.progsurf.2019.100540} {\bibfield  {journal}
  {\bibinfo  {journal} {Prog. Surf. Sci.}\ }\textbf {\bibinfo {volume} {94}},\
  \bibinfo {pages} {100540} (\bibinfo {year} {2019})}\BibitemShut {NoStop}%
\bibitem [{\citenamefont {Kuzmanovic}\ \emph {et~al.}(2020)\citenamefont
  {Kuzmanovic}, \citenamefont {Wu}, \citenamefont {Weideneder}, \citenamefont
  {Quay},\ and\ \citenamefont {Aprili}}]{kuzmanovic2020}%
  \BibitemOpen
  \bibfield  {author} {\bibinfo {author} {\bibfnamefont {M.}~\bibnamefont
  {Kuzmanovic}}, \bibinfo {author} {\bibfnamefont {B.~Y.}\ \bibnamefont {Wu}},
  \bibinfo {author} {\bibfnamefont {M.}~\bibnamefont {Weideneder}}, \bibinfo
  {author} {\bibfnamefont {C.~H.~L.}\ \bibnamefont {Quay}},\ and\ \bibinfo
  {author} {\bibfnamefont {M.}~\bibnamefont {Aprili}},\ }\bibfield  {title}
  {\bibinfo {title} {Evidence for spin-dependent energy transport in a
  superconductor},\ }\href {https://doi.org/doi.org/10.1038/s41467-020-18161-w}
  {\bibfield  {journal} {\bibinfo  {journal} {Nat. Commun.}\ }\textbf {\bibinfo
  {volume} {11}},\ \bibinfo {pages} {2041} (\bibinfo {year}
  {2020})}\BibitemShut {NoStop}%
\bibitem [{\citenamefont {Machon}\ \emph {et~al.}(2013)\citenamefont {Machon},
  \citenamefont {Eschrig},\ and\ \citenamefont {Belzig}}]{machon2013}%
  \BibitemOpen
  \bibfield  {author} {\bibinfo {author} {\bibfnamefont {P.}~\bibnamefont
  {Machon}}, \bibinfo {author} {\bibfnamefont {M.}~\bibnamefont {Eschrig}},\
  and\ \bibinfo {author} {\bibfnamefont {W.}~\bibnamefont {Belzig}},\
  }\bibfield  {title} {\bibinfo {title} {Nonlocal {{Thermoelectric Effects}}
  and {{Nonlocal Onsager}} relations in a {{Three-Terminal Proximity-Coupled
  Superconductor-Ferromagnet Device}}},\ }\href
  {https://doi.org/10.1103/PhysRevLett.110.047002} {\bibfield  {journal}
  {\bibinfo  {journal} {Phys. Rev. Lett.}\ }\textbf {\bibinfo {volume} {110}},\
  \bibinfo {pages} {047002} (\bibinfo {year} {2013})}\BibitemShut {NoStop}%
\bibitem [{\citenamefont {Ozaeta}\ \emph {et~al.}(2014)\citenamefont {Ozaeta},
  \citenamefont {Virtanen}, \citenamefont {Bergeret},\ and\ \citenamefont
  {Heikkil{\"a}}}]{ozaeta2014}%
  \BibitemOpen
  \bibfield  {author} {\bibinfo {author} {\bibfnamefont {A.}~\bibnamefont
  {Ozaeta}}, \bibinfo {author} {\bibfnamefont {P.}~\bibnamefont {Virtanen}},
  \bibinfo {author} {\bibfnamefont {F.~S.}\ \bibnamefont {Bergeret}},\ and\
  \bibinfo {author} {\bibfnamefont {T.~T.}\ \bibnamefont {Heikkil{\"a}}},\
  }\bibfield  {title} {\bibinfo {title} {Predicted {{Very Large Thermoelectric
  Effect}} in {{Ferromagnet-Superconductor Junctions}} in the {{Presence}} of a
  {{Spin-Splitting Magnetic Field}}},\ }\href
  {https://doi.org/10.1103/PhysRevLett.112.057001} {\bibfield  {journal}
  {\bibinfo  {journal} {Phys. Rev. Lett.}\ }\textbf {\bibinfo {volume} {112}},\
  \bibinfo {pages} {057001} (\bibinfo {year} {2014})}\BibitemShut {NoStop}%
\bibitem [{\citenamefont {Kolenda}\ \emph {et~al.}(2016)\citenamefont
  {Kolenda}, \citenamefont {Wolf},\ and\ \citenamefont
  {Beckmann}}]{kolenda2016}%
  \BibitemOpen
  \bibfield  {author} {\bibinfo {author} {\bibfnamefont {S.}~\bibnamefont
  {Kolenda}}, \bibinfo {author} {\bibfnamefont {M.~J.}\ \bibnamefont {Wolf}},\
  and\ \bibinfo {author} {\bibfnamefont {D.}~\bibnamefont {Beckmann}},\
  }\bibfield  {title} {\bibinfo {title} {Observation of {{Thermoelectric
  Currents}} in {{High-Field Superconductor-Ferromagnet Tunnel Junctions}}},\
  }\href {https://doi.org/10.1103/PhysRevLett.116.097001} {\bibfield  {journal}
  {\bibinfo  {journal} {Phys. Rev. Lett.}\ }\textbf {\bibinfo {volume} {116}},\
  \bibinfo {pages} {097001} (\bibinfo {year} {2016})}\BibitemShut {NoStop}%
\bibitem [{\citenamefont {Heidrich}\ and\ \citenamefont
  {Beckmann}(2019)}]{heidrich2019}%
  \BibitemOpen
  \bibfield  {author} {\bibinfo {author} {\bibfnamefont {J.}~\bibnamefont
  {Heidrich}}\ and\ \bibinfo {author} {\bibfnamefont {D.}~\bibnamefont
  {Beckmann}},\ }\bibfield  {title} {\bibinfo {title} {Nonlocal thermoelectric
  effects in high-field superconductor-ferromagnet hybrid structures},\ }\href
  {https://doi.org/10.1103/PhysRevB.100.134501} {\bibfield  {journal} {\bibinfo
   {journal} {Phys. Rev. B}\ }\textbf {\bibinfo {volume} {100}},\ \bibinfo
  {pages} {134501} (\bibinfo {year} {2019})}\BibitemShut {NoStop}%
\bibitem [{\citenamefont {Aikebaier}\ \emph {et~al.}(2018)\citenamefont
  {Aikebaier}, \citenamefont {Silaev},\ and\ \citenamefont
  {Heikkil{\"a}}}]{aikebaier2018}%
  \BibitemOpen
  \bibfield  {author} {\bibinfo {author} {\bibfnamefont {F.}~\bibnamefont
  {Aikebaier}}, \bibinfo {author} {\bibfnamefont {M.~A.}\ \bibnamefont
  {Silaev}},\ and\ \bibinfo {author} {\bibfnamefont {T.~T.}\ \bibnamefont
  {Heikkil{\"a}}},\ }\bibfield  {title} {\bibinfo {title} {Supercurrent-induced
  charge-spin conversion in spin-split superconductors},\ }\href
  {https://doi.org/10.1103/PhysRevB.98.024516} {\bibfield  {journal} {\bibinfo
  {journal} {Phys. Rev. B}\ }\textbf {\bibinfo {volume} {98}},\ \bibinfo
  {pages} {024516} (\bibinfo {year} {2018})}\BibitemShut {NoStop}%
\bibitem [{\citenamefont {H{\"u}bler}\ \emph {et~al.}(2010)\citenamefont
  {H{\"u}bler}, \citenamefont {Camirand~Lemyre}, \citenamefont {Beckmann},\
  and\ \citenamefont {{v. L{\"o}hneysen}}}]{huebler2010}%
  \BibitemOpen
  \bibfield  {author} {\bibinfo {author} {\bibfnamefont {F.}~\bibnamefont
  {H{\"u}bler}}, \bibinfo {author} {\bibfnamefont {J.}~\bibnamefont
  {Camirand~Lemyre}}, \bibinfo {author} {\bibfnamefont {D.}~\bibnamefont
  {Beckmann}},\ and\ \bibinfo {author} {\bibfnamefont {H.}~\bibnamefont {{v.
  L{\"o}hneysen}}},\ }\bibfield  {title} {\bibinfo {title} {Charge imbalance in
  superconductors in the low-temperature limit},\ }\href
  {https://doi.org/10.1103/PhysRevB.81.184524} {\bibfield  {journal} {\bibinfo
  {journal} {Phys. Rev. B}\ }\textbf {\bibinfo {volume} {81}},\ \bibinfo
  {pages} {184524} (\bibinfo {year} {2010})}\BibitemShut {NoStop}%
\bibitem [{\citenamefont {Wolf}\ \emph {et~al.}(2013)\citenamefont {Wolf},
  \citenamefont {H{\"u}bler}, \citenamefont {Kolenda}, \citenamefont
  {v.~L{\"o}hneysen},\ and\ \citenamefont {Beckmann}}]{wolf2013}%
  \BibitemOpen
  \bibfield  {author} {\bibinfo {author} {\bibfnamefont {M.~J.}\ \bibnamefont
  {Wolf}}, \bibinfo {author} {\bibfnamefont {F.}~\bibnamefont {H{\"u}bler}},
  \bibinfo {author} {\bibfnamefont {S.}~\bibnamefont {Kolenda}}, \bibinfo
  {author} {\bibfnamefont {H.}~\bibnamefont {v.~L{\"o}hneysen}},\ and\ \bibinfo
  {author} {\bibfnamefont {D.}~\bibnamefont {Beckmann}},\ }\bibfield  {title}
  {\bibinfo {title} {Spin injection from a normal metal into a mesoscopic
  superconductor},\ }\href {https://doi.org/10.1103/PhysRevB.87.024517}
  {\bibfield  {journal} {\bibinfo  {journal} {Phys. Rev. B}\ }\textbf {\bibinfo
  {volume} {87}},\ \bibinfo {pages} {024517} (\bibinfo {year}
  {2013})}\BibitemShut {NoStop}%
\bibitem [{\citenamefont {Wolf}\ \emph {et~al.}(2014)\citenamefont {Wolf},
  \citenamefont {S{\"u}rgers}, \citenamefont {Fischer},\ and\ \citenamefont
  {Beckmann}}]{wolf2014c}%
  \BibitemOpen
  \bibfield  {author} {\bibinfo {author} {\bibfnamefont {M.~J.}\ \bibnamefont
  {Wolf}}, \bibinfo {author} {\bibfnamefont {C.}~\bibnamefont {S{\"u}rgers}},
  \bibinfo {author} {\bibfnamefont {G.}~\bibnamefont {Fischer}},\ and\ \bibinfo
  {author} {\bibfnamefont {D.}~\bibnamefont {Beckmann}},\ }\bibfield  {title}
  {\bibinfo {title} {Spin-polarized quasiparticle transport in exchange-split
  superconducting aluminum on europium sulfide},\ }\href
  {https://doi.org/10.1103/PhysRevB.90.144509} {\bibfield  {journal} {\bibinfo
  {journal} {Phys. Rev. B}\ }\textbf {\bibinfo {volume} {90}},\ \bibinfo
  {pages} {144509} (\bibinfo {year} {2014})}\BibitemShut {NoStop}%
\bibitem [{\citenamefont {Giazotto}\ \emph {et~al.}(2006)\citenamefont
  {Giazotto}, \citenamefont {Heikkil{\"a}}, \citenamefont {Luukanen},
  \citenamefont {Savin},\ and\ \citenamefont {Pekola}}]{giazotto2006}%
  \BibitemOpen
  \bibfield  {author} {\bibinfo {author} {\bibfnamefont {F.}~\bibnamefont
  {Giazotto}}, \bibinfo {author} {\bibfnamefont {T.~T.}\ \bibnamefont
  {Heikkil{\"a}}}, \bibinfo {author} {\bibfnamefont {A.}~\bibnamefont
  {Luukanen}}, \bibinfo {author} {\bibfnamefont {A.~M.}\ \bibnamefont
  {Savin}},\ and\ \bibinfo {author} {\bibfnamefont {J.~P.}\ \bibnamefont
  {Pekola}},\ }\bibfield  {title} {\bibinfo {title} {Opportunities for
  mesoscopics in thermometry and refrigeration: Physics and applications},\
  }\href {https://doi.org/10.1103/RevModPhys.78.217} {\bibfield  {journal}
  {\bibinfo  {journal} {Rev. Mod. Phys.}\ }\textbf {\bibinfo {volume} {78}},\
  \bibinfo {pages} {217} (\bibinfo {year} {2006})}\BibitemShut {NoStop}%
\bibitem [{\citenamefont {Baselmans}\ \emph {et~al.}(1999)\citenamefont
  {Baselmans}, \citenamefont {Morpurgo}, \citenamefont {{van Wees}},\ and\
  \citenamefont {Klapwijk}}]{baselmans1999}%
  \BibitemOpen
  \bibfield  {author} {\bibinfo {author} {\bibfnamefont {J.~J.~A.}\
  \bibnamefont {Baselmans}}, \bibinfo {author} {\bibfnamefont {A.~F.}\
  \bibnamefont {Morpurgo}}, \bibinfo {author} {\bibfnamefont {B.~J.}\
  \bibnamefont {{van Wees}}},\ and\ \bibinfo {author} {\bibfnamefont {T.~M.}\
  \bibnamefont {Klapwijk}},\ }\bibfield  {title} {\bibinfo {title} {Reversing
  the direction of the supercurrent in a controllable {{Josephson}} junction},\
  }\href {https://doi.org/10.1038/16204} {\bibfield  {journal} {\bibinfo
  {journal} {Nature}\ }\textbf {\bibinfo {volume} {397}},\ \bibinfo {pages}
  {43} (\bibinfo {year} {1999})}\BibitemShut {NoStop}%
\bibitem [{\citenamefont {Maki}(1964)}]{maki1964}%
  \BibitemOpen
  \bibfield  {author} {\bibinfo {author} {\bibfnamefont {K.}~\bibnamefont
  {Maki}},\ }\bibfield  {title} {\bibinfo {title} {{Pauli Paramagnetism and
  Superconducting State. II}},\ }\href {https://doi.org/10.1143/PTP.32.29}
  {\bibfield  {journal} {\bibinfo  {journal} {Prog. Theor. Phys.}\ }\textbf
  {\bibinfo {volume} {32}},\ \bibinfo {pages} {29} (\bibinfo {year}
  {1964})}\BibitemShut {NoStop}%
\bibitem [{\citenamefont {Anthore}\ \emph {et~al.}(2003)\citenamefont
  {Anthore}, \citenamefont {Pothier},\ and\ \citenamefont
  {Esteve}}]{anthore2003}%
  \BibitemOpen
  \bibfield  {author} {\bibinfo {author} {\bibfnamefont {A.}~\bibnamefont
  {Anthore}}, \bibinfo {author} {\bibfnamefont {H.}~\bibnamefont {Pothier}},\
  and\ \bibinfo {author} {\bibfnamefont {D.}~\bibnamefont {Esteve}},\
  }\bibfield  {title} {\bibinfo {title} {Density of {{States}} in a
  {{Superconductor Carrying}} a {{Supercurrent}}},\ }\href
  {https://doi.org/10.1103/PhysRevLett.90.127001} {\bibfield  {journal}
  {\bibinfo  {journal} {Phys. Rev. Lett.}\ }\textbf {\bibinfo {volume} {90}},\
  \bibinfo {pages} {127001} (\bibinfo {year} {2003})}\BibitemShut {NoStop}%
\bibitem [{\citenamefont {Maki}(1969)}]{maki1969}%
  \BibitemOpen
  \bibfield  {author} {\bibinfo {author} {\bibfnamefont {K.}~\bibnamefont
  {Maki}},\ }\bibfield  {title} {\bibinfo {title} {Gapless
  {{Superconductivity}}},\ }in\ \href@noop {} {\emph {\bibinfo {booktitle}
  {Superconductivity}}},\ Vol.~\bibinfo {volume} {2},\ \bibinfo {editor}
  {edited by\ \bibinfo {editor} {\bibfnamefont {R.~D.}\ \bibnamefont {Parks}}}\
  (\bibinfo  {publisher} {{Dekker}},\ \bibinfo {address} {{New York}},\
  \bibinfo {year} {1969})\ p.\ \bibinfo {pages} {1035}\BibitemShut {NoStop}%
\bibitem [{\citenamefont {Alexander}\ \emph {et~al.}(1985)\citenamefont
  {Alexander}, \citenamefont {Orlando}, \citenamefont {Rainer},\ and\
  \citenamefont {Tedrow}}]{alexander1985}%
  \BibitemOpen
  \bibfield  {author} {\bibinfo {author} {\bibfnamefont {J.~A.~X.}\
  \bibnamefont {Alexander}}, \bibinfo {author} {\bibfnamefont {T.~P.}\
  \bibnamefont {Orlando}}, \bibinfo {author} {\bibfnamefont {D.}~\bibnamefont
  {Rainer}},\ and\ \bibinfo {author} {\bibfnamefont {P.~M.}\ \bibnamefont
  {Tedrow}},\ }\bibfield  {title} {\bibinfo {title} {Theory of {{Fermi-liquid}}
  effects in high-field tunneling},\ }\href
  {https://doi.org/10.1103/PhysRevB.31.5811} {\bibfield  {journal} {\bibinfo
  {journal} {Phys. Rev. B}\ }\textbf {\bibinfo {volume} {31}},\ \bibinfo
  {pages} {5811} (\bibinfo {year} {1985})}\BibitemShut {NoStop}%
\bibitem [{\citenamefont {Catelani}\ \emph {et~al.}(2008)\citenamefont
  {Catelani}, \citenamefont {Wu},\ and\ \citenamefont {Adams}}]{catelani2008}%
  \BibitemOpen
  \bibfield  {author} {\bibinfo {author} {\bibfnamefont {G.}~\bibnamefont
  {Catelani}}, \bibinfo {author} {\bibfnamefont {X.~S.}\ \bibnamefont {Wu}},\
  and\ \bibinfo {author} {\bibfnamefont {P.~W.}\ \bibnamefont {Adams}},\
  }\bibfield  {title} {\bibinfo {title} {Fermi-liquid effects in the gapless
  state of marginally thin superconducting films},\ }\href
  {https://doi.org/10.1103/PhysRevB.78.104515} {\bibfield  {journal} {\bibinfo
  {journal} {Phys. Rev. B}\ }\textbf {\bibinfo {volume} {78}},\ \bibinfo
  {pages} {104515} (\bibinfo {year} {2008})}\BibitemShut {NoStop}%
\bibitem [{\citenamefont {Bobkova}\ and\ \citenamefont
  {Bobkov}(2016)}]{bobkova2016}%
  \BibitemOpen
  \bibfield  {author} {\bibinfo {author} {\bibfnamefont {I.~V.}\ \bibnamefont
  {Bobkova}}\ and\ \bibinfo {author} {\bibfnamefont {A.~M.}\ \bibnamefont
  {Bobkov}},\ }\bibfield  {title} {\bibinfo {title} {Injection of
  nonequilibrium quasiparticles into {{Zeeman-split}} superconductors: A way to
  create long-range spin imbalance},\ }\href
  {https://doi.org/10.1103/PhysRevB.93.024513} {\bibfield  {journal} {\bibinfo
  {journal} {Phys. Rev. B}\ }\textbf {\bibinfo {volume} {93}},\ \bibinfo
  {pages} {024513} (\bibinfo {year} {2016})}\BibitemShut {NoStop}%
\end{thebibliography}%

\end{document}